\documentclass[journal]{IEEEtranTIE}
\usepackage{graphicx}
\usepackage{cite}
\usepackage{picinpar}
\usepackage{amsmath}
\usepackage{url}
\usepackage[latin1]{inputenc}
\usepackage{colortbl}
\usepackage{soul}
\usepackage{multirow}
\usepackage{pifont}
\usepackage{color}
\usepackage{alltt}
\usepackage[hidelinks]{hyperref}
\usepackage{enumerate}
\usepackage{siunitx}
\usepackage{breakurl}
\usepackage{epstopdf}
\usepackage{pbox}
\usepackage{svg}
\usepackage{emf}
\usepackage{lipsum}
\usepackage{soul}

\usepackage{amssymb}
\usepackage{amsfonts, amssymb}

\usepackage{balance}

\begin{document}
\title{A Highly Linear and Flexible FPGA-Based Time-to-Digital Converter  \\ }

\author{
	\vskip 1em
	
	Yuanyuan Hua, and Danial Chitnis 

	\thanks{
	
		Manuscript received July 27, 2021; revised October 16, 2021; accepted November 03, 2021.
		This work was supported in part by QuantIC project (https://quantic.ac.uk/) under EPSRC Grant EP/T00097X/1, and in part by National Natural Science Foundation of China (NSFC) under Grant 11603080.
		
		Yuanyuan Hua is with the Purple Mountain Observatory, Chinese Academy of Science, Nanjing, 210000, China, and the School of Engineering, Institute for Integrated Micro and Nano Systems, University of Edinburgh, Edinburgh, EH9 3FF, UK. 
		Danial Chitnis is with the School of Engineering, Institute for Integrated Micro and Nano Systems, University of Edinburgh, Edinburgh, EH9 3FF, UK.
		
        (corresponding author: Danial Chitnis;  e-mail: d.chitnis@ed.ac.uk).
	}
}
\maketitle
	
\begin{abstract}
Time-to-Digital Converters (TDCs) are major components for the measurements of time intervals. Recent developments in Field-Programmable Gate Array (FPGA) have enabled the opportunity to implement high-performance TDCs, which were only possible using dedicated hardware. In order to eliminate empty histogram bins and achieve a higher level of linearity, FPGA-based TDCs typically apply compensation methods either using multiple delay lines consuming more resources or post-processing, leading to a permanent loss of temporal information. We propose a novel TDC with a single delay line and without compensation to realize a highly linear TDC by encoding the states of the delay lines instead of the thermometer code used in the conventional TDCs. The experimental results show our states-based approach achieves an improved Differential Non-Linearity (DNL) of \mbox{[-0.998, -1.533]} for time resolution of 5.00~ps, \mbox{[-0.44,0.49]} for 10.04~ps, \mbox{[-0.16, 0.19]} for 21.65~ps, \mbox{[-0.10, 0.11]} for 43.87~ps, \mbox{[-0.06, 0.07]} for 64.12~ps, and \mbox{[-0.07, 0.05]} for 87.73~ps, whilst no empty bins have been observed. To our knowledge, the achieved raw linearity together with the zero empty bins and a simple delay line structure exceeds previously reported of the FPGA-based TDCs.  

%{\st{Experimental results show that the empty histogram bins are reduced to less than 0.1\% at the time resolution of 5.00ps, and have not been observed in the selected time resolutions of 10.04ps, 21.65ps, 43.87ps, 64.11ps, and 87.73ps.}}

\end{abstract}


\markboth{IEEE TRANSACTIONS ON INDUSTRIAL ELECTRONICS}%
{}

\definecolor{limegreen}{rgb}{0.2, 0.8, 0.2}
\definecolor{forestgreen}{rgb}{0.13, 0.55, 0.13}
\definecolor{greenhtml}{rgb}{0.0, 0.5, 0.0}

\section{Introduction}

\IEEEPARstart{T}{ime}-to-digital converters (TDCs) are devices which can measure the  time-related intervals with sub-100ps time resolution \cite{kalisz2003review }. They are vital components for applications which require a high time resolution such as Light Detection and Ranging (LiDAR) \cite{xie2021128,hejazi2020low, kim20217}, three-dimensional (3D) imaging \cite{hutchings2019reconfigurable,gyongy2020high }, Positron Emission Tomography (PET) \cite{shen20151, fishburn201319}, Fluorescence Lifetime Imaging (FLIM) \cite{tancock2019review}, Diffuse Optical Tomography \cite{lyons2019computational}, high energy physics	\cite{shen20151, wang2015nonlinearity}, and space exploration \cite{zhang2019high }.
A typical TDC is realized through either an analogue or a digital approach. The conventional analogue approach consists of two methods time: time-stretching and time-to-amplitude conversion \cite{kalisz2003review }. However, both methods require high-speed analogue-to-digital converters (ADCs) to digitize the analogue input signals. These ADCs are limited by their gain-bandwidth and aperture jitter from the process technology and introduce the inevitable quantization and thermal noises during the conversion stage of the TDC \cite{mahjoubfar2017time}. Generally, digital TDCs are less sensitive to process technology and temperature than their analogue counterparts \cite{won2015dual, kalisz2003review} due to the robustness to analog noise and signal drift. A digital TDC can be constructed with a Vernier oscillator or a delay line \cite{roberts2010brief, favi200917ps,fishburn201319, lai2017cost, dudek2000high, hwang2004high}. Vernier TDCs utilize the difference between the phase of their oscillators to measure the intervals, which are limited by their conversion rate due to numerous clock cycles for one measurement \cite{roberts2010brief}. A digital TDC with a delay-line makes the conversion at every clock cycle, leading to higher efficiency than the Vernier method.

Delay line based TDCs can be implemented in Application-Specific Integrated Circuits (ASICs) or Field Programmable Gate Arrays (FPGAs). An ASIC-based TDC requires dedicated implementation, which increases design iteration cycles and reduces the flexibility for the TDC to be utilized in various instrument configurations \cite{fishburn201319, hutchings2019reconfigurable, kim20149, al2018cmos, rehman201816, veerappan2011160, dutton201511}. FPGA-based TDC provides an alternative approach with a shorter development cycle for different requirements \cite{won2015dual, roberts2010brief, sui20182, zhang20177}. Recent developments in FPGA have also enabled the opportunity to implement high-performance digital TDCs. Vernier delay lines (VDLs) and Tapped Delay Lines (TDLs) are the main architectures used in the Digital TDC \cite{kim20149, song2006high, garzetti2021time}.  TDL-based TDCs consist of a simpler structure with a faster conversion rate than VDL-based TDCs, making them more popular in recent studies\cite{chen2018multichannel, kim20149}.

There has been a continuous development of FPGA-based TDL-TDCs in the past two decades. In 1997, a direct time-to-digital converter based on tapped delay lines was proposed. The achieved resolution was 200~ps with the Differential Non-Linearity (DNL) of [-0.47, 0.47] \cite{kalisz1997field}. In 2000, a TDC with the Least Significant Bit (LSB) of 110~ps and DNL of 1.88~LSB was published \cite{szplet2000interpolating }. In 2006, a TDC with Look-up Table (LUT) as delay elements with a resolution of 65.8~ps and DNL within [-0.953, 1.051] \cite{song2006high }. In 2009, a 17~ps TDC using carry logic with DNL [-1, 3.55] was reported \cite{favi200917ps }. In 2013, a TDC with the sub-20~ps resolution and a DNL of [-1, 1.5] was implemented \cite{ fishburn201319}. In 2016, a dual-phase TDL-TDC was reported, and the effects of the clock skew were discussed in detail \cite{won2015dual}. In 2019, the impact of the manufacturing process on the root mean square (RMS) of a sub-TDL topology was studied \cite{chen2018multichannel}. 

\begin{figure}[!t]\centering
%\begin{figure}[htbp]
  \includegraphics[width=8.5cm]{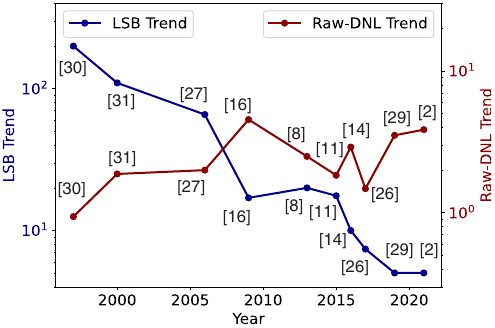}
  \caption{Time resolution and raw DNL trend within the past 20 years, showing LSB (time resolution) decreasing over time while raw peek-to-peak DNL which is the DNL before compensation has not improved.}
  \label{fig:tdc_history}
  
\end{figure}

The twenty years' development, as shown in Fig. \ref{fig:tdc_history}, indicates a clear improvement in the resolution from 200~ps to sub-10~ps. However, the raw linearity of TDC, which is the linearity of raw data prior to any compensation method, has not improved over time. Multichain topology and post-processing are the two main methods previously used for the optimization of linearity. Multichain topology uses parallel delay lines to obtain an average of the bins \cite{won2015dual, sui20182, andaloussi2002novel}. This method is able to achieve small and uniform bins, however, it increases the complexity and leads to higher utilization of hardware resources. The post-processing method, which has appeared in recent studies, calibrate and compensate for the raw data to improve Differential Non-Linearity (DNL) and Integral Non-Linearity (INL) \cite{chen2018multichannel, xie2021128, choi2020design, jarvinen2021100}. However, due to the random variations in the clock skew and process mismatch, the misplacement of time intervals in the raw TDL output remains unpredictable, leading to a permanent loss of temporal information.

\begin{figure}[!t]\centering
%\begin{figure}[htbp]
      \includegraphics[width=8.5cm]{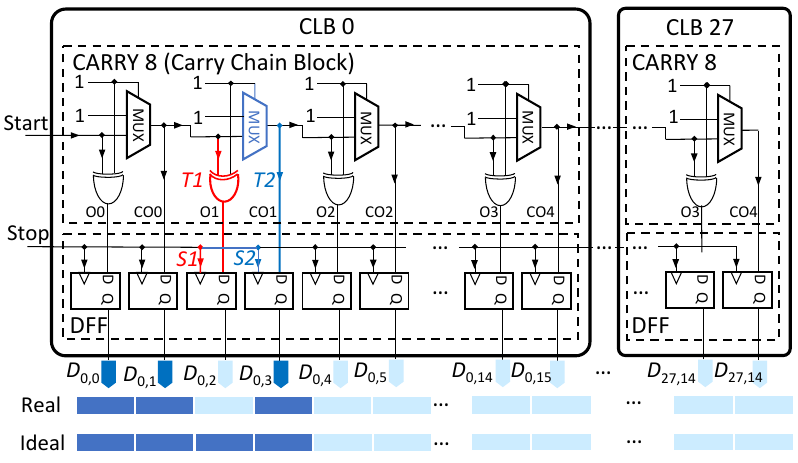}
  \caption{Block diagram of Configurable Logic Block (CLB) in Xilinx UltraScale+ MPSoC FPGA, and an example of ideal and real thermometer code.}
  \label{fig:CLB_diagram}
\end{figure}

A traditional TDL-based TDC uses the thermometer code to encode time intervals. In an ideal delay line, there is no gap between the codes. However, in a real delay line, there may be one or more gaps, as shown in Fig. \ref{fig:CLB_diagram}. In this figure, $T1$, $T2$, $S1$, $S2$ represent the propagation delay of each route. If a Stop edge occurs in the proximity of $D_{0,3}$, then ideally, $D_{0,0}$ to $D_{0,3}$ are expected to have the same logic states while $D_{0,4}$ to $D_{0,15}$ have the opposite logic state.  However, when $T1< T2$ and $S1=S2$, then $D_{0,2}$ will have the same logic state as $D_{0,4}$ to $D_{0,15}$, hence, a gap appears in the real code resulting to a difference from the ideal code. The gap leads to a bubble error in traditional TDCs which use the ideal thermometer code. The variations of $T1$, $T2$, $S1$, and $S2$ are mainly due to the effects of the clock skew and process mismatch, which are unavoidable in electronic systems. 

The output of the delay units representing the combined results of all the delay effects, including clock skew and process mismatch, are referred to as real states, which are the states of the internal carry chain. We propose a novel TDC which interprets and sequences the real states from the TDL and encodes the TDC with the real states instead of using the ideal thermometer codes. This method minimizes the multiple effects of clock skew and mismatch on the TDL and contributes to a predictable and flexible raw linearity. Proof-of-concept experiments demonstrate that the state-based approach achieves a high raw-linearity and significantly reduces the empty histogram bins while using only a simple delay line.

%%%%%%%%%%%%%%%%%%%%%%%%%%%%%%%%%%%%%%%%%%%%%%%%%

\section{PROPOSED METHODOLOGY}  

\begin{figure}[!t]\centering
%\begin{figure}[htbp]
  \includegraphics[width=8.5 cm]{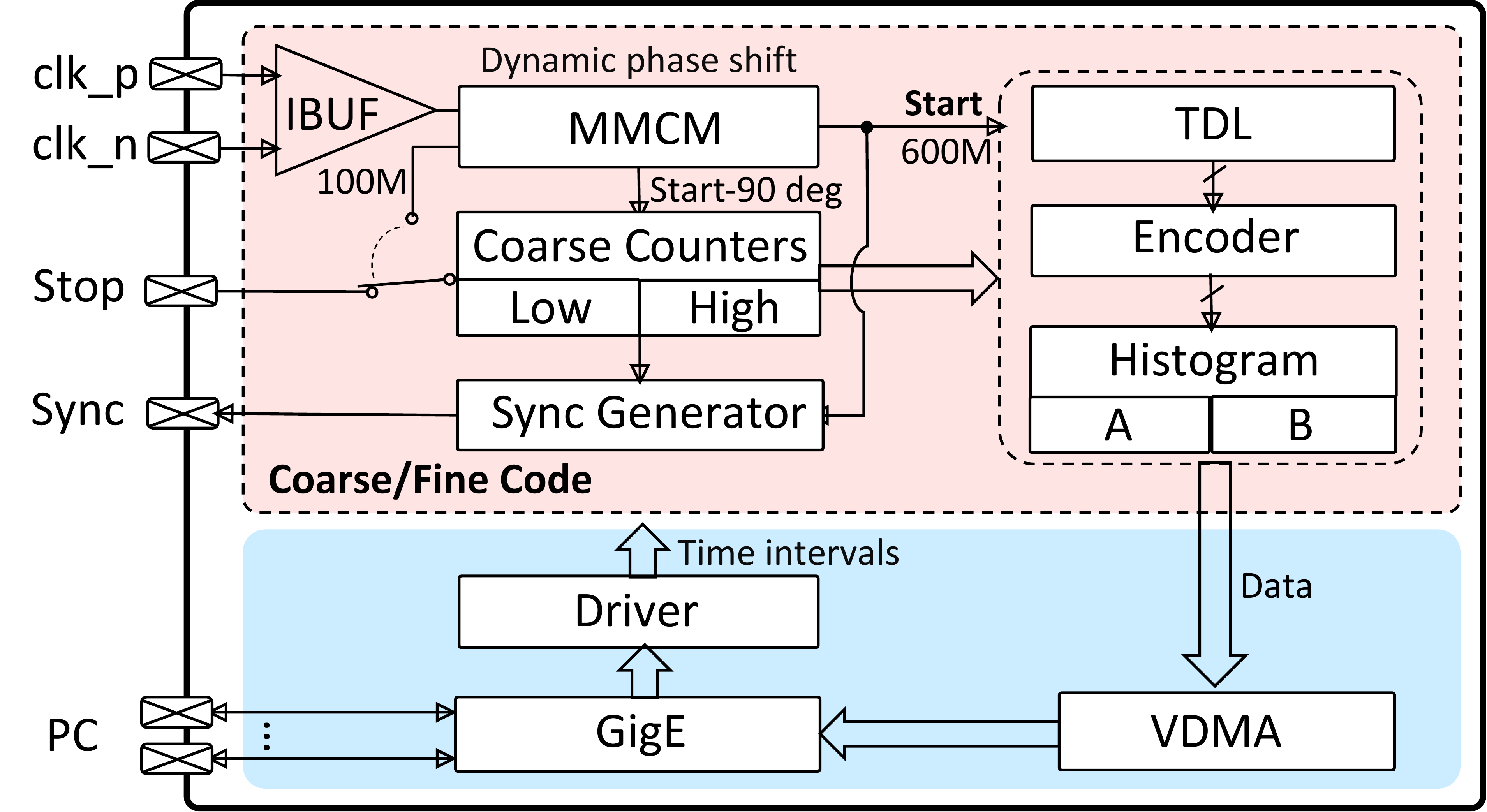}
  \caption{Block diagram of the proposed flexible TDC. The programmable logic (in red) and processing system (in blue).}
  \label{fig:TDC_diagram}
\end{figure}

A 16 nm Xilinx UltraScale+ MPSoC FPGA is utilized as the platform to implement the TDC and demonstrate the efficiency of the states-based approach. The block diagram of the TDC is shown in Fig. \ref{fig:TDC_diagram}. A Mixed-Mode Clock Manager (MMCM) is applied to provide a 600 MHz clock as the Start signal for the TDL. The 100 MHz Stop signal used in the time interval measurement is also generated from the MMCM, which is externally controlled by a Personal Computer (PC) through a Gigabit network and an embedded driver within the processor part of the FPGA. Random signals generated from a Single-Photon Avalanche Diode (SPAD)\cite{richardson2009low} are used as the Stop signal in the code density test. The TDC consists of both coarse code and fine code. The coarse code counts the cycles of the Start signal while each cycle covers the measurement range of the fine code. Two interleaved histograms are generated on-chip and transferred by a Gigabit network through Video Direct Memory Access (VDMA). The main abbreviation and nomenclature applied in this paper are shown in TABLE \ref{table_1}.

\subsection{The states-based TDC}
\label{sq:state-based-TDC}

\begin{figure}[!t]\centering
%\begin{figure}[htbp]
  \includegraphics[width=8.5cm]{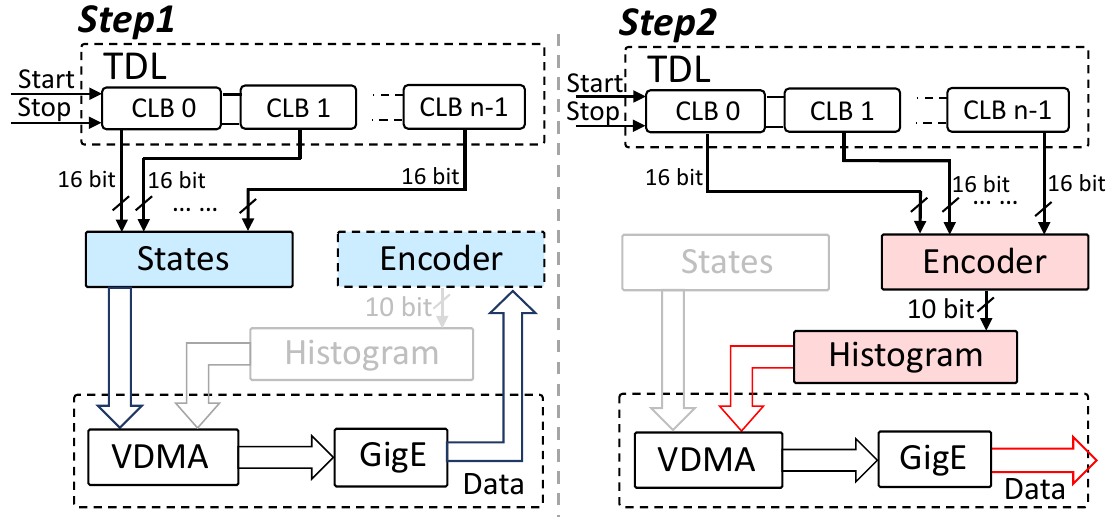}
  \caption{The two steps for the collection and application of the states. (a) Step 1: states collection and modification of the Encoder.  (b) Step 2: histogram generation and data transfer.}
  \label{fig:step12}
\end{figure}

\begin{table}[!t]
	\renewcommand{\arraystretch}{1.3}
	\caption{Nomenclature and Abbreviation}
	\centering
	\label{table_1}
	%\centering
	\resizebox{\columnwidth}{!}{
		\begin{tabular}{l l l}
			\hline\hline \\[-3mm]
			Nomenclature & \hphantom{spaces} & Referred to as   \\[1.6ex] \hline
			TDC  &  \hphantom{spaces} & Time to Digital Converters  \\
			TDL  &  \hphantom{spaces} & Tapped Delay Line  \\
			LSB  &  \hphantom{spaces} & Least Significant Bit  \\
			DNL  &  \hphantom{spaces} & Differential Non-Linearity  \\
		    INL  &  \hphantom{spaces} & Integral Non-Linearity \\
			FPGA  &  \hphantom{spaces} & Field Programmable Gate Array  \\
			ASIC  &  \hphantom{spaces} & Application-Specific Integrated Circuit\\
			MMCM  &  \hphantom{spaces} & Mixed-Mode Clock Manager  \\
			DFF  &  \hphantom{spaces} & D Flip Flop  \\
			CLB  &  \hphantom{spaces} & Configurable Logic Block  \\
			
			$RSE$  &  \hphantom{spaces} & Relative Standard Error  \\
			$N$    &  \hphantom{spaces} & Total number of configured groups  \\
			$D_{c,b}$  &  \hphantom{spaces} & Bit value of CLBs (bit: $b+16\times c$)  \\
			$s[j]$  &  \hphantom{spaces} & Width of the $j$-th state (time delay)  \\
			$ref$  &  \hphantom{spaces} & Expected time resolution  \\
			$W[i]$  &  \hphantom{spaces} & Width of the $i$-th group (time delay)  \\
			$C_i$  &  \hphantom{spaces} & Count number of $i$-th group (Code density test)  \\
			$Seq$  &  \hphantom{spaces}& Sum of all bit value of CLBs  \\
			$FP$  &  \hphantom{spaces} & First Pass in bin configuration  \\
			$SP$ &  \hphantom{spaces} & Second Pass in bin configuration \\  

			\hline\hline
		\end{tabular}
	}
\end{table}

The proposed TDC is constructed using Configurable Logic Blocks (CLBs) which are the most widely available resources in the Xilinx UltraScale+ MPSoC FPGAs. Each CLB is composed of a carry chain with multiplexers, XOR gates and D-Flip-Flops (DFF). The traditional TDL-based TDC applies the ideal thermometer codes for encoding and assumes all gates in the CLBs are identical. Additionally, it assumes that the DFFs sample propagation states simultaneously. However, as Fig. \ref{fig:CLB_diagram} illustrates, the real states contain unavoidable gaps which vary with the ideal thermometer code. The generated timestamps from the ideal code cannot represent the actual time, which leads to reduced linearity.  
\begin{figure}[!t]\centering
%\begin{figure}[htbp]
  \includegraphics[width=8.5cm]{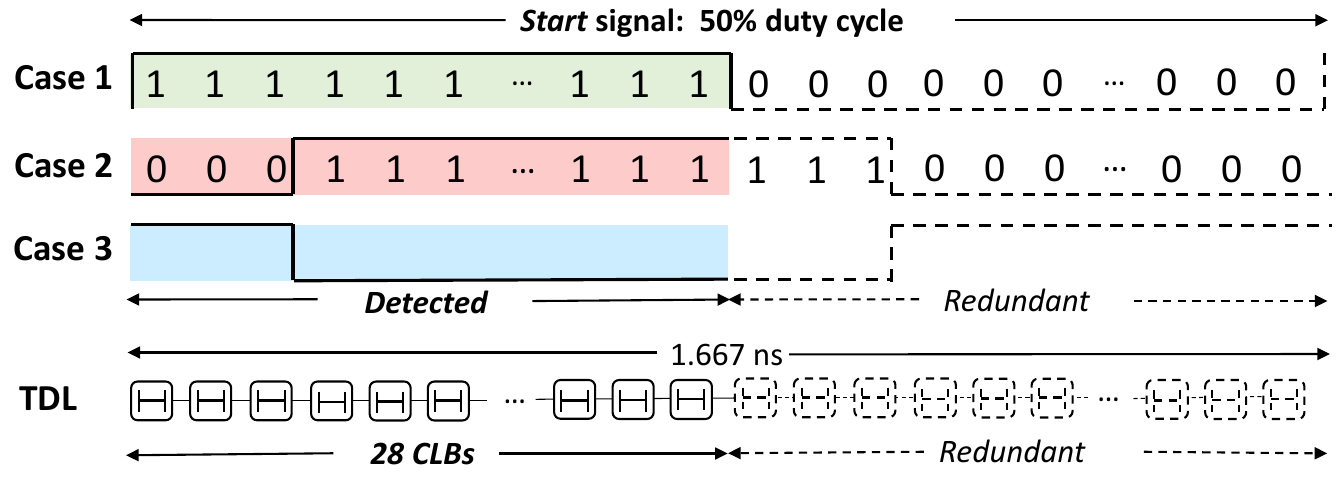}
  \caption{A demonstration of half-sized TDL with three typical example cases.}
  \label{fig:halfsize}
\end{figure} 
In this work, all the blocks in Fig.\ref{fig:TDC_diagram} are based on a Xilinx Zynq UltraScale+ xczu3cg-sfvc784-1-e FPGA. 
In the implementation, real states which are the outputs of the CLBs are collected and used to encode the timestamps. As the Start signal (in Fig. \ref{fig:CLB_diagram} ) propagates through the CLB, the DFF registers each state with a binary code. Each real state is considered as a bin, while the relative propagation of each state is the bin width. Fig. \ref{fig:step12} shows the two steps required to collect and apply the real states. In step 1, all states were collected directly from the TDL, while an iterative algorithm, shown in (\ref{eq:Seqi}), was applied to sequence these states.  In (\ref{eq:Seqi}), $D_{c,b}$ is the logic state of the CLB as shown Fig. \ref{fig:CLB_diagram}, while $b$ is the position of $D_{c,b}$ in the CLB, $c$ is the position of the CLB in the delay line, and 28 CLBs are used in total in this work. Each state has a $Seq$ value which is the reference for the alignment of the states. 
In cases when two states have the same $Seq$ according to (\ref{eq:Seqi}), for 5~ps time resolution, we chose to keep the two states independently to ensure high resolution and sort them according to the latest 1~(High) to 0~(Low), or 0~(Low) to 1~(High) position within the state. In other time resolutions (10.04~ps, 21.65~ps, etc.), the adjacent states with same $Seq$ are combined together to improve the time interval measurement. These sequenced states were then applied to the Encoder to generate timestamps, as shown in Fig. \ref{fig:step12}(a). In step 2, the TDL was connected back to the updated Encoder for the histogram generation and data output. 

The code density tests are performed with the real states for the bin width calculations. Although these tests can reveal the empty histogram bins, they do not reveal the missing codes, both of which contribute to the non-linearity.
A missing code is defined as the missing output of the CLB, which are not recorded by the Encoder, hence, they do not appear in the histogram.
The missing code occurs when the TDC input does not have a matched real state in the Encoder, hence, the time input is not recorded and may lead to an blank histogram. 
The code density test is performed by applying uniformly random pulses to the TDC as Stop signals. The Encoder inside the TDC includes the output of the CLBs, which represents the states of the Stop signals. It encodes these states into unique numbers and then writes them to the histograms. If a state was not included in the Encoder, it could not be recorded in the histogram, which means that state is missing.
The time interval measurement is a method to identify missing codes by searching for the blank histograms, which contain all zero bins. Furthermore, the time interval measurements verify the sequential order of the real states and the functionality of the TDC. 

\begin{align}
\label{eq:Seqi}
Seq=\sum_{c=0}^{M-1} \sum_{b=0}^{15} D_{c,b} \
\end{align}

\begin{figure*}[!t]\centering
%\begin{figure}[htbp]
  \includegraphics[width=18cm]{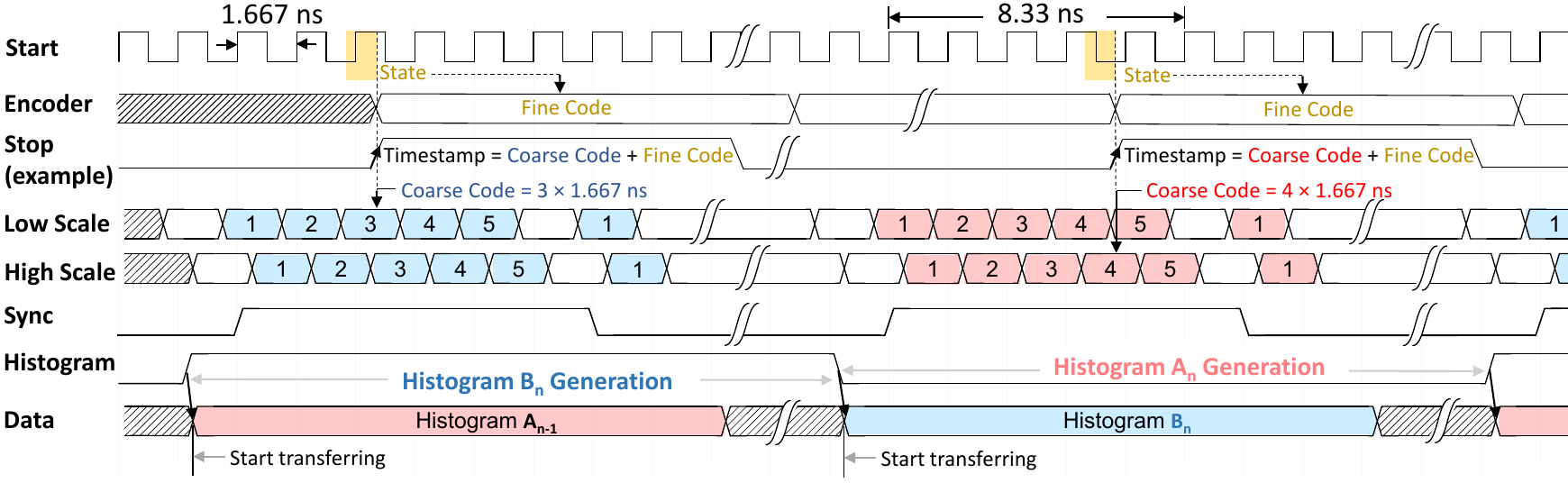}
  \caption{A timing diagram of the on-chip histogram generator, showing the timestamp calculation and the continuous operation of the interleaved histogram generation without deadtime. The Low Scale and High Scale are coarse counters with a 90-degree phase difference to avoid the timing race condition. The  duration of the histograms are set by the integration time of the TDC. }
  \label{fig:digital}
\end{figure*}

\subsection{Half-sized TDL and histogram generator}

A 600 MHz signal with a duty cycle of 50\% is generated by the MMCM as the Start signal of the TDC. When a Stop event occurs, the output of the CLBs, which is the sampled Start signal, is considered the Fine Code. The Low Scale and High Scale are interleaved coarse counters to extend the measurement range, based on the fine code measurement range of 1.667~ns. The 90-degree phase offset of the coarse counters with the Start signal is used to avoid a timing race condition. 
Typically, one cycle of the Start signal equals the propagation delay of the TDL. 
However, if the Start signal has a duty cycle of 50\% and half of the signal detected, the remaining half is redundant to be detected.

\begin{figure}[!t]\centering
%\begin{figure}[htbp]
  \includegraphics[width=8.5cm]{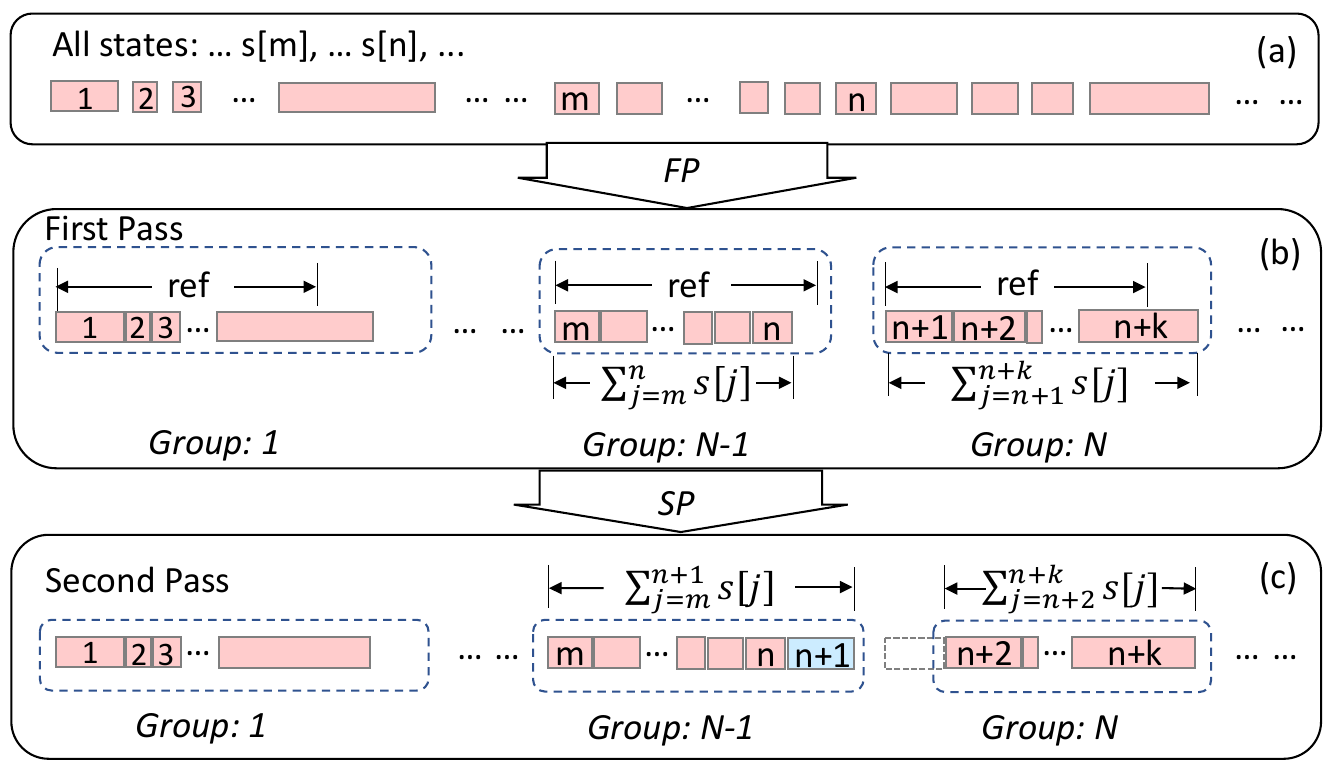}
  \caption{A demonstration of the two-pass bin configuration method. (a) bin widths of the real states are prepared for a two-pass bin configuration. (b) the first pass for coarse configuration. (c) the second pass for fine configuration.}
  \label{fig:flowchart}
\end{figure} 

Under this consideration, we shortened the TDL to be half cycle of the Start signal. Fig. \ref{fig:halfsize} is an ideal example demonstrating how a half-sized TDL samples every possible Start signal and generates the states. With the half-sized TDL, which is 833.5~ps in length, the example Start signals such as Case 1, Case 2, and Case 3 shown in Fig. \ref{fig:halfsize} can be uniquely recorded without the requirement of the other half part, which leads to a 50\% reduction of the resources for the delay line.

An on-chip histogram generator was applied, with histogram A and histogram B in an interleaved configuration for continuous data collection and transfer. As shown in Fig. \ref{fig:digital}, the duration of the histograms is set by the integration time of the TDC which can be changed based on user requirements. The interleaving of histograms A and B eliminates the deadtime and ensures a continuous time-to-digital conversion. Each histogram includes 1200 bins with 16 bits depth. The Sync is the output trigger which is used to synchronize with the external devices such as a laser driver in a LiDAR system. 

\begin{figure}[!t]\centering
%\begin{figure}[htbp]
  \includegraphics[width=8.8cm]{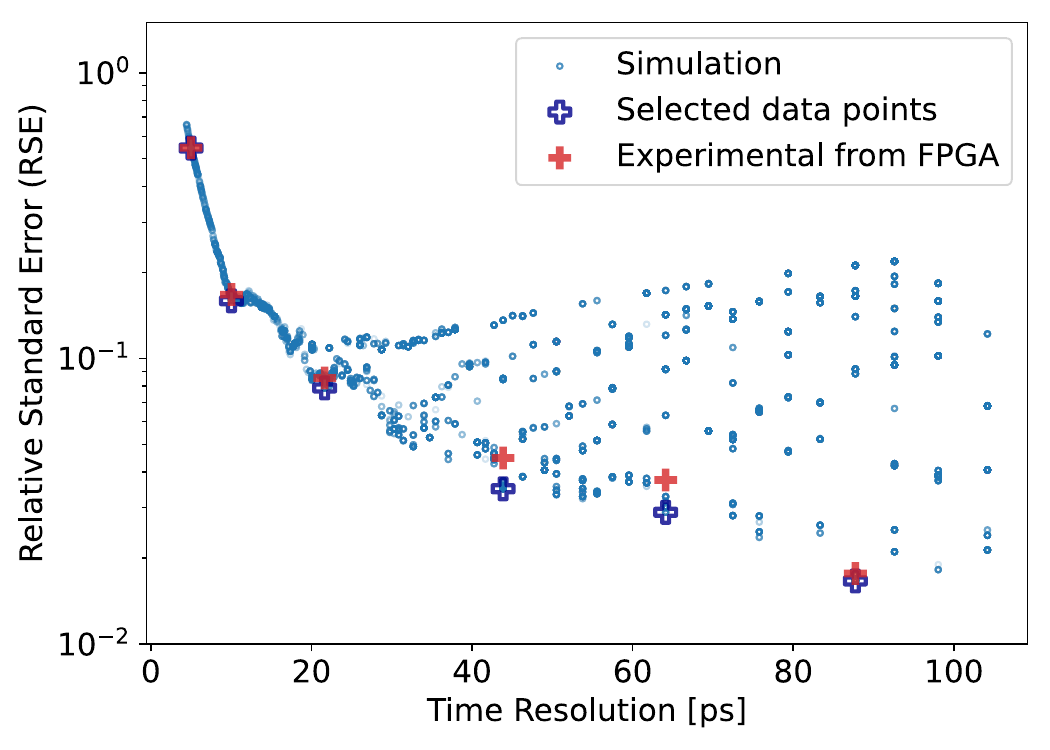}
  \caption{Relative Standard Error ($RSE$) versus time resolution for each bin configuration after applying the two-pass method, demonstrating an incremental decrease in the $RSE$ with increasing time-resolution. The blue \textcolor{blue}{$\circ$} is simulation results from 19.6k individual bin configuration, the blue hollow \textbf{\textcolor{blue}{+}}  is the selected time resolution for experiments, and the red \textbf{\textcolor{red}{+}} are the experimental results after FPGA implementation.}
  \label{fig:RSE}
\end{figure}

%\begin{figure*}[!t]
\begin{figure*}[htbp]\centering

  \includegraphics[width=18.3cm]{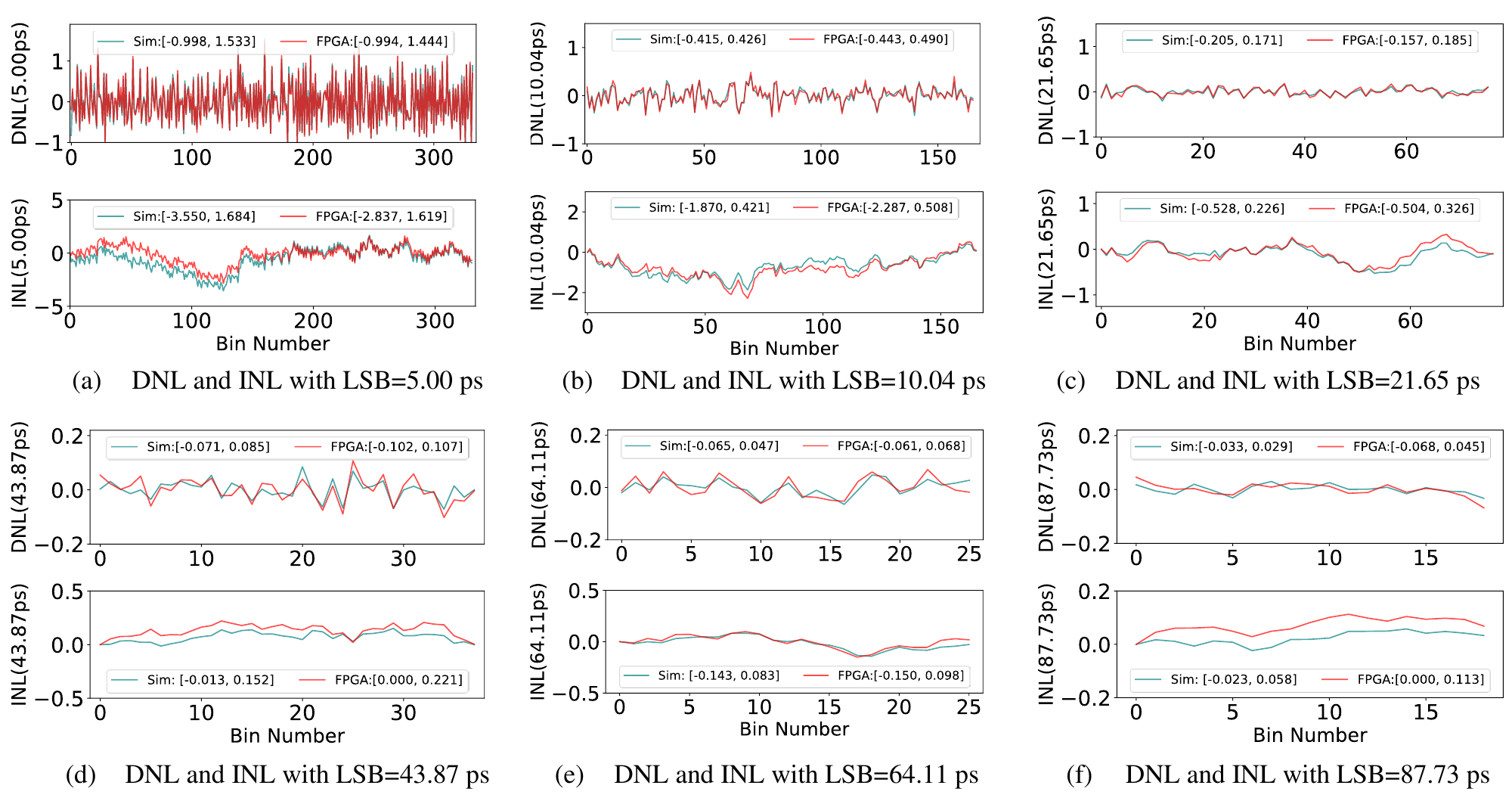}

  \caption{Simulation and experimental DNL and INL results, with a measurement range of 1.667~ns. (a) LSB=5~ps. (b) LSB=10.04~ps. (c) LSB=21.65~ps. (d) LSB =43.87~ps. (e) LSB =64.11~ps. (f) LSB =87.73~ps.   }\label{fig:DNLINL}
\end{figure*}

\subsection{Relative Standard Error-based bin configuration}  \label{Relative Standard Error-based bin configuration}

The propagation delay of each state represents the minimum bin widths which are calculated from the code density test \cite{chen2018multichannel}. The bin widths can be combined for different time resolution requirements. The process is referred to as bin configuration. As shown in Fig. \ref{fig:flowchart}, the bin widths of the real states are prepared for a two-pass bin configuration as $s[m]$ or $s[n]$, while $m$ or $n$ is the sequence of the state and $0<m<n$. 
%{\st{The First Pass (FP) is the coarse configuration, while the Second Pass (SP) is the fine configuration.} }
The $ref$ corresponds to the desired time resolution, providing a reference for the combinations of the first pass, which coarsely categorizes the bins into groups. If $FP$ of (\ref{eq:first}) is positive, the bins such as $s[m]$ to $s[n]$ would be combined as one group while $s[n+1]$ starts the following combination.

\begin{align}
\label{eq:first}
FP=|ref-\sum_{j=m}^{n+1}s[j]|-|ref-\sum_{j=m}^{n}s[j]|
\end{align}

In the second pass, the first bin in each group is moved to the previous group for the fine configuration. The calculation was performed according to (\ref{eq:second}). If $SP$ is positive, then the first bin in each group such as $s[n+1]$ would be attached to the group ahead, while $s[n+2]$ would be selected as the starting bin for the following combination with the value of $SP$ in (\ref{eq:second}).

\begin{multline}
\label{eq:second}
          SP=|\sum_{i=m}^{n}s[i]-\sum_{i=n+1}^{n+k}s[i]|
-|\sum_{i=m}^{n+1}s[i]-\sum_{i=n+2}^{n+k}s[i]|  
\end{multline}

Hence, the Second Pass, $SP$, increase the accuracy of the First Pass.
The metric for evaluating the configurations after the second pass is the Relative Standard Error ($RSE$). As shown in (\ref{eq:RSE}) and Fig. \ref{fig:flowchart}, $N$ is the total number of groups, while $W[i]$ and $\bar{W}$ are the individual and average width of the groups, respectively. The average width of the groups is referred to as the time resolution of the TDC, which is also referred to as Least Significant Bit (LSB).

Each $RSE$ corresponds to a bin configuration, while the bin configuration with the least $RSE$ is chosen to simulate the DNL and INL. Based on the simulation, the bin configuration can be written to the encoder of the TDC for real DNL and INL calculation. This approach leads to a fast and reliable method to predict linearity before implementation.

\begin{figure}[!t]\centering
%\begin{figure}[htbp]
  \includegraphics[width=8.8 cm]{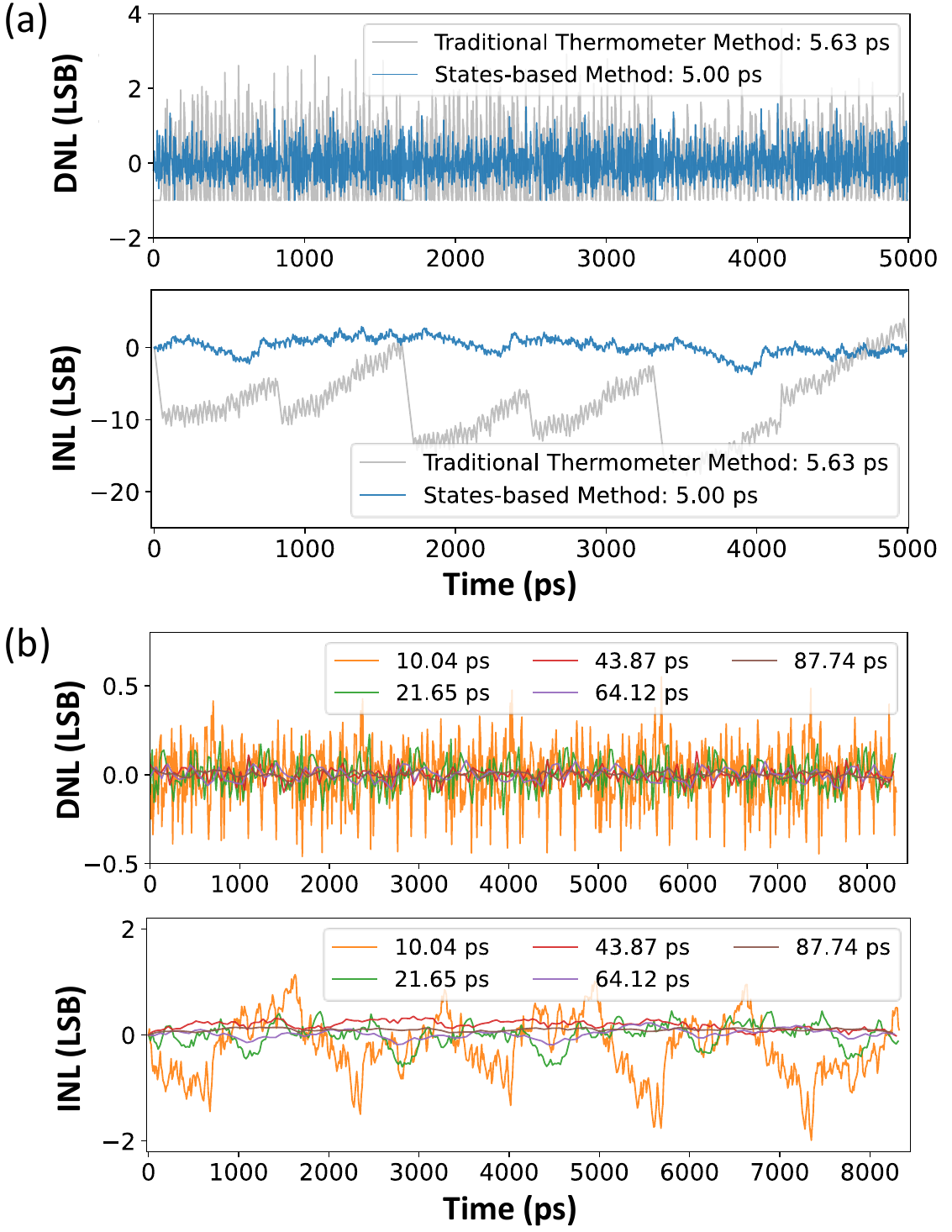}
  \caption{DNL and INL measurement results with the measurement range of (a) 5~ns at 5~ps time resolution, and the traditional thermometer method testing result with 5.63 ps time resolution. (b) the measurement range of 8~ns for the rest time resolution with the states-based method.}
  \label{fig:longDNLINL}
\end{figure}

\begin{figure}[!t]\centering
%\begin{figure}[htbp]
  \includegraphics[width=8.8 cm]{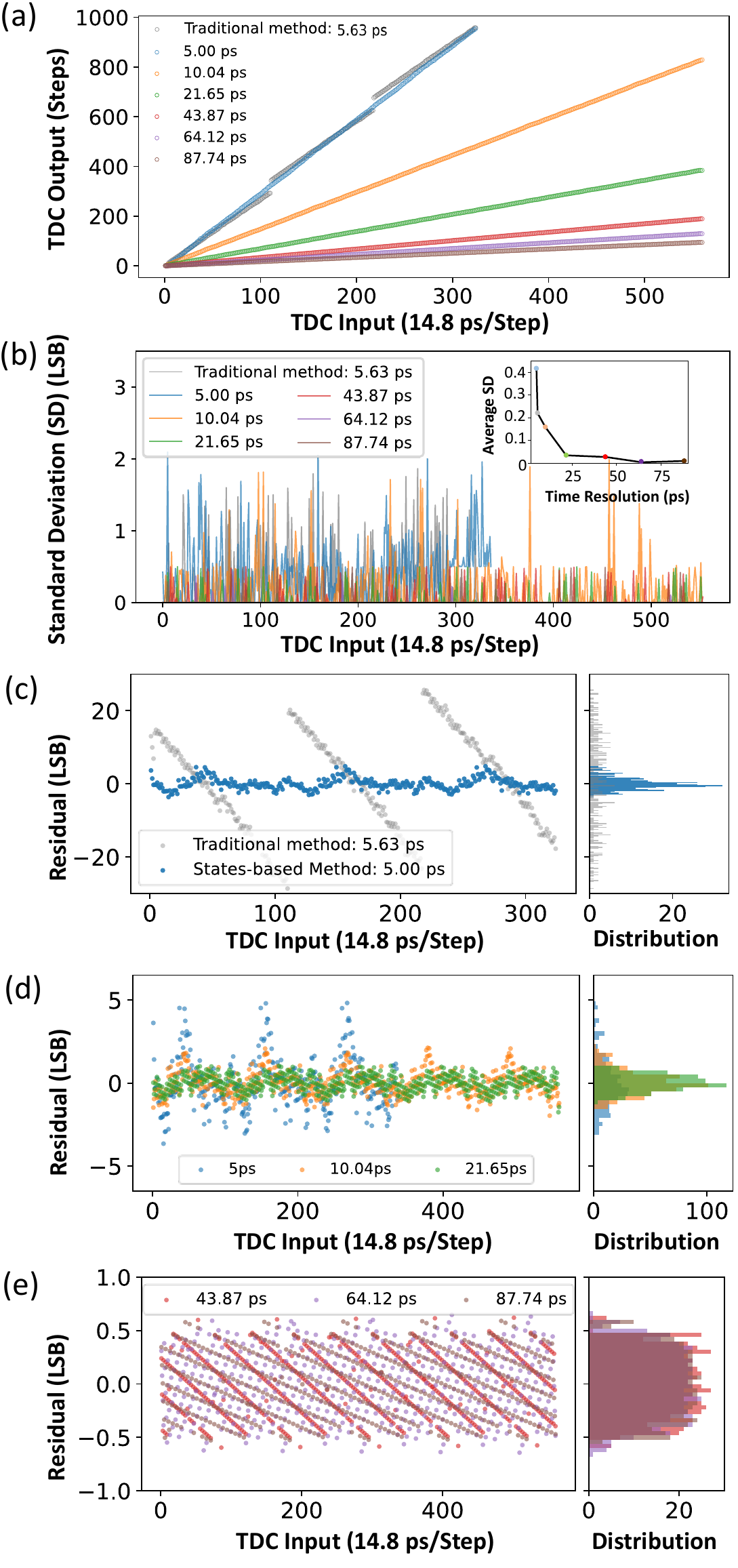}
  \caption{The results of the time interval measurements for the selected time resolution. (a) The TDC output (position of the highest bin in the histogram) versus TDC input (delay steps), including the testing result of the 5.63 time resolution with the traditional thermometer method. (b) The Standard Deviation of each dot in (a). (c) The residuals of 5.00~ps and 5.63~ps with traditional method. (d) The residuals of 10.04~ps, and 21.65~ps time resolution and the residuals' distribution. (e) The residuals of 43.87~ps, 64.12~ps, and 87.74~ps time resolution and the residuals' distribution.}
  \label{fig:longtime}
\end{figure}

\begin{align}
\label{eq:RSE}
RSE=\sqrt{\frac{1}{N-1} \sum_{i=1}^{N} (W[i]-\bar{W})^2} \: / \: \bar{W}
\end{align}

%%%%%%%%%%%%%%%%%%%%%%%%%%%%%%%%%%%%%%%%%

\section{EXPERIMENTS AND RESULTS}

The efficiency of the proposed approach is presented in this section, which includes the experiments and results of RSE-based bin configuration, code density test with both fine and coarse code, and long-term time interval measurement. A Single-Photon Avalanche Diode (SPAD) sensor is used as the event source to provide random signals for the code density test\cite{richardson2009low}.

\subsection{Bin configuration}

The reason for using RSE as a metric for the characterization of the TDC is to determine the best bin configuration, which leads to the highest linearity. The bin widths of the states used for the RSE simulation are obtained from the earlier code density test. According to the analysis in the RSE-based bin configuration described in Section \ref{Relative Standard Error-based bin configuration}, the bin configuration starts from the value set of $ref$, representing the desired bin width which is then incrementally increased from 2 to 100~ps in the simulation.
Each $ref$ values are applied in $FP$ using (\ref{eq:first}), and the resulting configuration is then used for $SP$ in (\ref{eq:second}). After $SP$, a bin configuration corresponding to a $ref$ value is finished and used for RSE calculations. The time resolution is calculated from the bin configuration using the total Group number.
The time resolution is calculated by $1.667\: ns/N$ where the $1.667\: ns$ is the fine code measurement range, and $N$ is the total number of the groups. Fig. \ref{fig:RSE} shows the simulation and experiment results of the relationship between RSE and time resolution of the TDC. 
%{\st{Each blue $\circ$ corresponds to an RSE with a specific bin configuration. Multiple bin configurations may have same $N$ leading to the same time resolution. As a result, a single time resolution may correspond to multiple RSEs, as shown in Fig.}}
Each blue $\circ$ represents a unique $ref$ value. In the First Pass, $FP$, the total Group number $N$ is determined by sweeping the $ref$ in (\ref{eq:first}) from 2~ps to 100~ps with a resolution of 0.01~ps. After the Second Pass, $SP$, different $ref$ values may result in the same total Group number $N$. The time resolution is uniquely determined by the total Group number $N$, however, each group may have a different combination of bin configurations, resulting in different $RSE$ values. As a result, multiple $ref$ values may lead to the same time resolution with different RSE values.

The bin configurations which has the smallest RSEs represents the best linearity for that time resolution.
The RSE is based on (\ref{eq:RSE}), which is the ratio between the standard error and the average bin width (i.e. the time resolution). When the time resolution decreases from 5~ps to 100~ps, both standard error and the time resolution increase. However, the lowest standard error does not increase at the same rate of the time resolution, hence, the lowest RSE improves with decreasing time resolution from 5~ps to 100~ps.

%\begin{align}
%\label{eq:timeresolution}
%Time\; Resolution= 1.667ns/N 
%\end{align}

In order to verify the accuracy of the RSE simulation, six representative time resolutions 5.00~ps, 10.04~ps, 21.65~ps, 43.87~ps, 64.11~ps, 87.73~ps from Fig. \ref{fig:RSE} were selected.  The experimental results show the selected time resolution from the simulation and the FPGA implementation, which have a relative variation within 15\%. Based on the bin configurations from the RSE simulation, code density test and time interval test were performed for the six representative time resolutions.

\subsection{Code density test}

Code density test is one of the primary methods to measure the linearity of the TDC. It uses a random signal source to achieve a uniformly distributed histogram to calculate DNLs and INLs. 
In this experiment, a SPAD is exposed to constant intensity to generate random pulses \cite{tawfeeq2009random}. The Start signals from an oscillator and the Stop signal from the SPAD contribute to a uniform distribution in the histogram. The random pulses from the SPAD, enable histogram frame rate to reach over 1000 frames per second with each frame over 10000 counts, which is efficient for the DNL and INL measurement.

The $DNL_{i}$ and $INL_{i}$ are calculated for each group of bin configurations using (\ref{eq:DNLINL}) where $C_{i}$ is the count number of the individual group, which represents the relative group width. The  $C_{Avg}$ equals to $\sum_{i=1}^{N}{C_{i}}/N$ where $N$ is the number of the groups.

%\[DNL_{i}= \frac{C_{i}-C_{Avg}}{C_{Avg}}\]

\begin{align}
\label{eq:DNLINL}
DNL_{i}= \frac{C_{i}-C_{Avg}}{C_{Avg}}\;\;\; and\;\;\;  INL_{i}= \sum_{j=1}^{i} DNL_{j} 
\end{align}

After the simulations of DNLs and INLs based on the bin configurations, the six selected time resolutions were implemented to the FPGA. The comparisons of the simulation and implementation results are shown in Fig. \ref{fig:DNLINL}. The implementation results are in close agreement with the simulations of DNL and INL, which both improve as the time resolution decrease. Fig. \ref{fig:DNLINL} show that the relations between the time resolution (LSB) versus the linearity follow a similar trend to the time resolution versus RSE.

Based on the fine code test results shown in Fig. \ref{fig:DNLINL}, the initial measurement range of 1.667~ns was expanded to 5~ns for the selected 5.00~ps time resolution and a range of 8~ns at the remaining time resolutions, while the histogram records a maximum of 1200 bins. As Fig. \ref{fig:longDNLINL} shows, the long-range (5~ns and 8~ns) DNL and INL maintain similar linearity as the short-range (1.667~ns). No empty histogram bins have been observed in all selected time resolutions including 5.00~ps, 10.04~ps, 21.65~ps, 43.87~ps, 64.11~ps, and 87.73~ps. Other than this, the traditional thermometer method were applied to make a comparison with the proposed method as shown in Fig. \ref{fig:longDNLINL} (a). with the time resolution of 5.63~ps, the DNL and INL of the traditional thermometer method is [-1.00, 3.60] and [-17.7, 3.98], respectively, with over 20\% bins empty, while the states-based method at 5.00~ps achieves the DNL and INL of [-0.99, 1.58] and [-3.74, 2.89], respectively, with no empty bins observed.

\subsection{Time interval measurement}

Time interval measurement is an essential method to determine the functionality of a TDC. In this measurement, small time-steps are generated between the Start and the Stop signals, which sweep the entire measurement range of the TDC. In this implementation, both TDC Start and Stop signals were generated inside the FPGA using the MMCM block. The interpolated fine phase shift (IFPS) of the MMCM was used to generate the small time-steps between the Start and the Stop signals with a minimum step size of 14.8~ps.

Of all the histograms collected in the measurement, more than 90\% have a Full Width at Half Maximum (FWHM) of less than two bins while the remaining histograms have less than four bins populated. For each step, the bin in the histogram with the highest count number was considered the TDC output. Time resolutions including 5.00~ps, 10.04~ps, 21.65~ps, 43.87~ps, 64.11~ps, 87.73~ps were measured with step-size of 14.8~ps.
Based on the DNL results in Fig.\ref{fig:DNLINL} and according to (\ref{eq:DNLINL}), we can calculate the smallest bin for each time resolution. For time resolutions of 21.65~ps, 43.87~ps, 64.11~ps, and 87.73~ps, the smallest bin is larger than 18.25~ps, hence, the phase shift step of 14.8~ps covers all the bins, including the smallest bin width. However, for time resolutions of 5.00~ps and 10.04~ps, the smallest bins are 0.03~ps and 5.60~ps, respectively, which are smaller than the phase shift step of 14.8~ps, hence, it is not possible to cover all the bins with the current technology. As the primary purpose of the time interval measurement is to confirm the linear trend between the TDC input and the TDC output, the 14.8~ps phase shift steps are sufficient to confirm the linear trends.

\begin{table*}[!t]
	\renewcommand{\arraystretch}{1.3}
	\caption{A comparison with recent FPGA/ASIC-based TDCs}
	\centering
	\label{table:comparasion}
\begin{tabular}{c}

   \includegraphics[width=18cm]{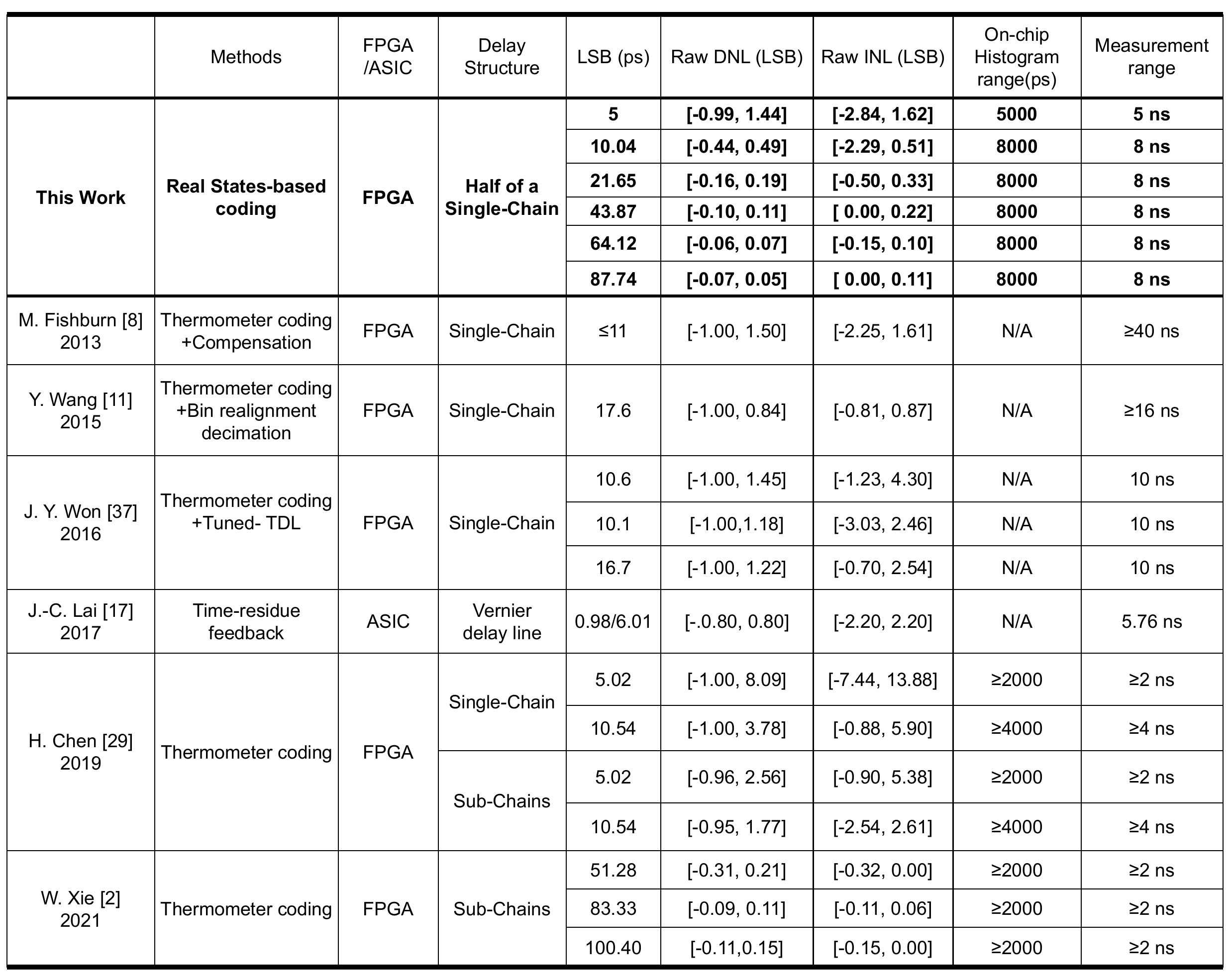}

\end{tabular}

\end{table*}

A linear regression model was applied to determine the linearity of the measurement results. The residuals of the linear regression model are calculated by the difference between the experimental TDC output from the predicated line at a given TDC input using $TDC_{output}- Predicted_{output}$. Each histogram has a measurement count of over 10k, and 1000 frames of histograms were saved for the average TDC output calculation, leading to approximately 10M total counts for each data point. The Standard Deviation of each TDC output, which is calculated from 1000 frames of histograms, improves as the time resolution decreases from 5.00~ps to 87.74~ps. No significant Standard Deviation differences occur between the traditional thermometer method and the proposed states-based method, which potentially indicates that the Standard Deviation was mainly affected by the jitters of the Start and Stop signal rather than the TDC itself.

%\begin{align}
%\label{eq:residuals}
%Residuals= TDC_{output}- Predicted_{output} 
%\end{align}

From Fig. \ref{fig:longtime}(c), the peak-to-peak residuals' range of the 5.00~ps time resolution is [-5.5, 6.92], while the 10.04~ps and 21.65~ps are [-1.99, 3.96] and [-0.86, 0.90] respectively. A similar trend occurs for the 43.96~ps, 64.12~ps and 87.74~ps bin widths which have a residual of  [-0.78, 0.70], [-0.65, 0.64], and [-0.58, 0.60], respectively.
As Fig. \ref{fig:longtime}(c), Fig. \ref{fig:longtime}(d), and Fig. \ref{fig:longtime}(e) show, the distribution of the residuals shares a similar trend as the DNL and INL, which achieve improvement as the time resolution decreases from 5.00~ps to 87.74 ps. The testing results of the traditional thermometer method at 5.63~ps are shown in Fig. \ref{fig:longtime}(a), Fig. \ref{fig:longtime}(b), and Fig. \ref{fig:longtime}(c), with decreased linear trend and more distributed residuals compared with the other states-based time resolutions. The FWHM of the residual distribution of 5.00 ps, 10.04 ps and 21.56 ps is 3.8 LSB, 2.2 LSB, and 1.4 LSB, respectively, while not much difference is shown on 43.87 ps, 64.12 ps, and 87.74 ps, which are approximately 1.2 LSB.     

The oscillations shown in Fig. \ref{fig:longtime}(c), Fig. \ref{fig:longtime}(d), \ref{fig:longtime}(e) are due to repetition of the fine code in the entire measurement range. Since the measurement range is 5~ns or 8~ns, and the fine measurement range is 1.667~ns, the fine code is repeated 3  or 5 times. As described in Section \ref{sq:state-based-TDC} no blank histograms were observed in Fig. \ref{fig:longtime}(a), hence, no missing code was found in this experiment.

%%%%%%%%%%%%%%%%%%%%%%%%%%%%%%%%%%%%%%%%%%%%%%%%%%

\section{Discussion}

The proposed state-based TDC is compared with few other approaches in TABLE \ref{table:comparasion} \cite{fishburn201319,wang2015nonlinearity,won2016time,lai2017cost, chen2018multichannel, xie2021128}. we have demonstrated our result in six separate time resolutions between 5 and 100~ps. The results from raw DNL measurement show a minimum DNL of above -1, which, according to (\ref{eq:DNLINL}), meaning the smallest bin is non-zero, hence, no empty bins are recorded. Additionally, the maximum raw DNL shows significant improvement relative to other TDCs with a single-chain or sub-chain delay structure. The Vernier structure of the delay line implemented in \cite{won2016time} may results in a better raw DNL relative to single-chain or sub-chain based TDCs, however, the Vernier structure is more complex and has slower frame rates.
Additionally, the proposed RSE-based bin configuration uses a simple RSE equation relative to the MB approach described in \cite{xie2021128}. On the other hand, Our approach achieves a better raw DNL at higher resolution such as 51.28~ps due to the proposed state-based concept rather than the traditional thermometer concept implemented in \cite{xie2021128}.

Moreover, we have achieved a high similarity between the simulation and experimental results in DNL, INL and RSE measurements, which provides a linearity reference before the time-consuming compiling and implementing of the code. Our implementation has a longer on-chip 8~ns full measurement range (5~ns full range for the 5.00~ps time resolution) relative to the previous implementations in TABLE \ref{table:comparasion}. The on-chip interleaved histogram generated in our implementation enables high frame rate outputs without deadtime for real-time applications.
Furthermore, our entire TDC system is implemented on a generic low-cost FPGA development board, enabling potential uses in applications with cost constraints, including robotics and Internet-of-Things (IoT). In order to make a fully embedded states-based TDC system without a PC, future developments may include the calculations of the $SP$, $FP$, and the linearity analysis of our RSE-based bin configuration inside the same FPGA.

%%%%%%%%%%%%%%%%%%%%%%%%%
 
\section{Conclusion}
We have developed a states-based TDC with improved raw linearity and flexible time resolution. The TDL consumes half of the traditional TDL resources whilst maintaining a continuous conversion without deadtime using interleaved histograms. The proposed RSE-based bin configuration approach is able to predict the combination of the states and inform the implementation of the TDC. We selected six different time resolutions for the verification of this approach. The RSE from the implementation compared to the simulation shows less than 15\% variation. The six different time resolutions were evaluated individually using code density measurements, while the implementation results were consistent with the simulated DNL and INL. This RSE-based approach provides a fast and reliable method to predict the linearity of a TDC prior to its implementation. Moreover, an extended measurement range of 8~ns (5~ns for 5.00~ps time resolution) was evaluated, with the experimental results reproducing the linearity of the shorter range of the fine code. Time interval measurements were taken for each selected time resolution to verify the functionality of the TDC. Using a linear regression model, the residuals of the TDC output for each time resolution were found to be in agreement with measured DNL results.

In summary, the main contributions of the proposed TDC are as follows:

\begin{enumerate}[1)]
	\item The concept of the states-based TDC is proposed. Compared with the traditional thermometer code based TDC, this method encodes the TDC with real states from the TDL, which contributes to the higher raw linearity and eliminates empty histogram bins.
    \item The RSE-based bin configuration is able to predict the time resolution, DNL and INL prior to the implementation hence achieving a TDC with flexible time resolution and the highest level of linearity at certain time resolution.
	\item A half-sized TDL, which utilizes half of the resources of one traditional delay line, is proposed. Additionally, interleaved histograms are applied for continuous data collection and transfer without deadtime.	
	\item Code density test and time interval test were applied to evaluate the efficiency of the proposed method for time resolutions from 5.00~ps to 87.74~ps. Results show the states based TDC achieves better raw linearity compared with the previous studies, with DNL of \mbox{[-1.00, -1.53]} for 5.00~ps, \mbox{[-0.44, 0.49]} for 10.04~ps and \mbox{[-0.07, 0.05]} for 87.73~ps, with no empty bins observed.

\end{enumerate}

Currently, only limited data were collected which may not cover all the potential states within the TDL. These states with rare occurrences are potentially missed, affecting the bin configuration and reducing linearity. Moreover, the states are potentially influenced by the temperature and variations of the Start signal of the TDC which is also the system's clock, where more tests need to be done\cite{pan201420}. In conclusion, the FPGA-based TDC provides a level of high linearity and flexibility to rapidly develop instrumentation in a large variety of time-of-flight applications including, LiDAR, 3D imaging, and healthcare monitoring.

\section*{Acknowledgment}
The authors thank Prof. Robert Henderson for influential suggestions and Dr. Alistair Gorman for helpful discussions.

\bibliographystyle{Bibliography/IEEEtranTIE}
\balance     % added by YY to sequence the references

\bibliography{Bibliography/IEEEabrv,Bibliography/Reference}\ %IEEEabrv instead of IEEEfull

\begin{thebibliography}{10}
\providecommand{\url}[1]{#1}
\csname url@samestyle\endcsname
\providecommand{\newblock}{\relax}
\providecommand{\bibinfo}[2]{#2}
\providecommand{\BIBentrySTDinterwordspacing}{\spaceskip=0pt\relax}
\providecommand{\BIBentryALTinterwordstretchfactor}{4}
\providecommand{\BIBentryALTinterwordspacing}{\spaceskip=\fontdimen2\font plus
\BIBentryALTinterwordstretchfactor\fontdimen3\font minus
  \fontdimen4\font\relax}
\providecommand{\BIBforeignlanguage}[2]{{%
\expandafter\ifx\csname l@#1\endcsname\relax
\typeout{** WARNING: IEEEtran.bst: No hyphenation pattern has been}%
\typeout{** loaded for the language `#1'. Using the pattern for}%
\typeout{** the default language instead.}%
\else
\language=\csname l@#1\endcsname
\fi
#2}}
\providecommand{\BIBdecl}{\relax}
\BIBdecl

\bibitem{kalisz2003review}
J.~Kalisz, ``Review of methods for time interval measurements with picosecond
  resolution,'' \emph{Metrologia}, vol.~41, no.~1, p.~17, 2003.

\bibitem{xie2021128}
W.~Xie, Y.~Wang, H.~Chen, and D.~D.-U. Li, ``{128-channel high-linearity
  resolution-adjustable time-to-digital converters for LiDAR applications:
  software predictions and hardware implementations},'' \emph{IEEE Transactions
  on Industrial Electronics}, 2021.

\bibitem{hejazi2020low}
A.~Hejazi, S.~Oh, M.~R.~U. Rehman, R.~E. Rad, S.~Kim, J.~Lee, Y.~Pu, K.~C.
  Hwang, Y.~Yang, and K.-Y. Lee, ``{A Low-Power Multichannel Time-to-Digital
  Converter Using All-Digital Nested Delay-Locked Loops With 50-ps Resolution
  and High Throughput for LiDAR Sensors},'' \emph{IEEE Transactions on
  Instrumentation and Measurement}, vol.~69, no.~11, pp. 9262--9271, 2020.

\bibitem{kim20217}
B.~Kim, S.~Park, J.-H. Chun, J.~Choi, and S.-J. Kim, ``{7.2 A 48$\times$ 4013.5
  mm Depth Resolution Flash LiDAR Sensor with In-Pixel Zoom Histogramming
  Time-to-Digital Converter},'' in \emph{2021 IEEE International Solid-State
  Circuits Conference (ISSCC)}, vol.~64, pp. 108--110.\hskip 1em plus 0.5em
  minus 0.4em\relax IEEE, 2021.

\bibitem{hutchings2019reconfigurable}
S.~W. Hutchings, N.~Johnston, I.~Gyongy, T.~Al~Abbas, N.~A. Dutton, M.~Tyler,
  S.~Chan, J.~Leach, and R.~K. Henderson, ``{A reconfigurable 3-D-stacked SPAD
  imager with in-pixel histogramming for flash LIDAR or high-speed
  time-of-flight imaging},'' \emph{IEEE Journal of Solid-State Circuits},
  vol.~54, no.~11, pp. 2947--2956, 2019.

\bibitem{gyongy2020high}
I.~Gyongy, S.~W. Hutchings, A.~Halimi, M.~Tyler, S.~Chan, F.~Zhu,
  S.~McLaughlin, R.~K. Henderson, and J.~Leach, ``{High-speed 3D sensing via
  hybrid-mode imaging and guided upsampling},'' \emph{Optica}, vol.~7, no.~10,
  pp. 1253--1260, 2020.

\bibitem{shen20151}
Q.~Shen, S.~Liu, B.~Qi, Q.~An, S.~Liao, P.~Shang, C.~Peng, and W.~Liu, ``{A 1.7
  ps equivalent bin size and 4.2 ps RMS FPGA TDC based on multichain
  measurements averaging method},'' \emph{IEEE Transactions on Nuclear
  Science}, vol.~62, no.~3, pp. 947--954, 2015.

\bibitem{fishburn201319}
M.~Fishburn, L.~H. Menninga, C.~Favi, and E.~Charbon, ``{A 19.6 ps, FPGA-based
  TDC with multiple channels for open source applications},'' \emph{IEEE
  transactions on nuclear science}, vol.~60, no.~3, pp. 2203--2208, 2013.

\bibitem{tancock2019review}
S.~Tancock, E.~Arabul, and N.~Dahnoun, ``A review of new time-to-digital
  conversion techniques,'' \emph{IEEE transactions on Instrumentation and
  Measurement}, vol.~68, no.~10, pp. 3406--3417, 2019.

\bibitem{lyons2019computational}
A.~Lyons, F.~Tonolini, A.~Boccolini, A.~Repetti, R.~Henderson, Y.~Wiaux, and
  D.~Faccio, ``Computational time-of-flight diffuse optical tomography,''
  \emph{Nature Photonics}, vol.~13, no.~8, pp. 575--579, 2019.

\bibitem{wang2015nonlinearity}
Y.~Wang and C.~Liu, ``{A nonlinearity minimization-oriented resource-saving
  time-to-digital converter implemented in a 28 nm Xilinx FPGA},'' \emph{IEEE
  Transactions on Nuclear Science}, vol.~62, no.~5, pp. 2003--2009, 2015.

\bibitem{zhang2019high}
Z.~Zhang, H.~Zhang, M.~Long, H.~Deng, Z.~Wu, and W.~Meng, ``{High precision
  space debris laser ranging with 4.2 W double-pulse picosecond laser at 1 kHz
  in 532nm},'' \emph{Optik}, vol. 179, pp. 691--699, 2019.

\bibitem{mahjoubfar2017time}
A.~Mahjoubfar, D.~V. Churkin, S.~Barland, N.~Broderick, S.~K. Turitsyn, and
  B.~Jalali, ``Time stretch and its applications,'' \emph{Nature Photonics},
  vol.~11, no.~6, pp. 341--351, 2017.

\bibitem{won2015dual}
J.~Y. Won, S.~I. Kwon, H.~S. Yoon, G.~B. Ko, J.-W. Son, and J.~S. Lee,
  ``{Dual-phase tapped-delay-line time-to-digital converter with on-the-fly
  calibration implemented in 40 nm FPGA},'' \emph{IEEE transactions on
  biomedical circuits and systems}, vol.~10, no.~1, pp. 231--242, 2015.

\bibitem{roberts2010brief}
G.~W. Roberts and M.~Ali-Bakhshian, ``A brief introduction to time-to-digital
  and digital-to-time converters,'' \emph{IEEE Transactions on Circuits and
  Systems II: Express Briefs}, vol.~57, no.~3, pp. 153--157, 2010.

\bibitem{favi200917ps}
C.~Favi and E.~Charbon, ``{A 17ps time-to-digital converter implemented in 65nm
  FPGA technology},'' in \emph{Proceedings of the ACM/SIGDA international
  symposium on Field programmable gate arrays}, pp. 113--120, 2009.

\bibitem{lai2017cost}
J.-C. Lai and T.-Y. Hsu, ``Cost-effective time-to-digital converter using
  time-residue feedback,'' \emph{IEEE Transactions on Industrial Electronics},
  vol.~64, no.~6, pp. 4690--4700, 2017.

\bibitem{dudek2000high}
P.~Dudek, S.~Szczepanski, and J.~V. Hatfield, ``{A high-resolution CMOS
  time-to-digital converter utilizing a Vernier delay line},'' \emph{IEEE
  Journal of Solid-State Circuits}, vol.~35, no.~2, pp. 240--247, 2000.

\bibitem{hwang2004high}
C.-S. Hwang, P.~Chen, and H.-W. Tsao, ``A high-precision time-to-digital
  converter using a two-level conversion scheme,'' \emph{IEEE Transactions on
  nuclear science}, vol.~51, no.~4, pp. 1349--1352, 2004.

\bibitem{kim20149}
K.~Kim, W.~Yu, and S.~Cho, ``{A 9 bit, 1.12 ps resolution 2.5 b/stage pipelined
  time-to-digital converter in 65 nm CMOS using time-register},'' \emph{IEEE
  Journal of Solid-State Circuits}, vol.~49, no.~4, pp. 1007--1016, 2014.

\bibitem{al2018cmos}
T.~Al~Abbas, N.~A. Dutton, O.~Almer, N.~Finlayson, F.~M. Della~Rocca, and
  R.~Henderson, ``{A CMOS SPAD sensor with a multi-event folded flash
  time-to-digital converter for ultra-fast optical transient capture},''
  \emph{IEEE Sensors Journal}, vol.~18, no.~8, pp. 3163--3173, 2018.

\bibitem{rehman201816}
S.~U. Rehman, M.~M. Khafaji, C.~Carta, and F.~Ellinger, ``{A 16 mW 250 ps
  double-hit-resolution input-sampled time-to-digital converter in 45-nm
  CMOS},'' \emph{IEEE Transactions on Circuits and Systems II: Express Briefs},
  vol.~65, no.~5, pp. 562--566, 2018.

\bibitem{veerappan2011160}
C.~Veerappan, J.~Richardson, R.~Walker, D.-U. Li, M.~W. Fishburn, Y.~Maruyama,
  D.~Stoppa, F.~Borghetti, M.~Gersbach, R.~K. Henderson \emph{et~al.}, ``A
  160$\times$ 128 single-photon image sensor with on-pixel 55ps 10b
  time-to-digital converter,'' in \emph{2011 IEEE International Solid-State
  Circuits Conference}, pp. 312--314.\hskip 1em plus 0.5em minus 0.4em\relax
  IEEE, 2011.

\bibitem{dutton201511}
N.~A. Dutton, S.~Gnecchi, L.~Parmesan, A.~J. Holmes, B.~Rae, L.~A. Grant, and
  R.~K. Henderson, ``{11.5 A time-correlated single-photon-counting sensor with
  14GS/S histogramming time-to-digital converter},'' in \emph{2015 IEEE
  International Solid-State Circuits Conference-(ISSCC) Digest of Technical
  Papers}, pp. 1--3.\hskip 1em plus 0.5em minus 0.4em\relax IEEE, 2015.

\bibitem{sui20182}
T.~Sui, Z.~Zhao, S.~Xie, Y.~Xie, Y.~Zhao, Q.~Huang, J.~Xu, and Q.~Peng, ``{A
  2.3-ps RMS resolution time-to-digital converter implemented in a low-cost
  cyclone V FPGA},'' \emph{IEEE transactions on instrumentation and
  measurement}, vol.~68, no.~10, pp. 3647--3660, 2018.

\bibitem{zhang20177}
M.~Zhang, H.~Wang, and Y.~Liu, ``{A 7.4 ps FPGA-based TDC with a 1024-unit
  measurement matrix},'' \emph{Sensors}, vol.~17, no.~4, p. 865, 2017.

\bibitem{song2006high}
J.~Song, Q.~An, and S.~Liu, ``A high-resolution time-to-digital converter
  implemented in field-programmable-gate-arrays,'' \emph{IEEE Transactions on
  Nuclear Science}, vol.~53, no.~1, pp. 236--241, 2006.

\bibitem{garzetti2021time}
F.~Garzetti, N.~Corna, N.~Lusardi, and A.~Geraci, ``{Time-to-Digital Converter
  IP-Core for FPGA at State of the Art},'' \emph{IEEE Access}, 2021.

\bibitem{chen2018multichannel}
H.~Chen and D.~D.-U. Li, ``{Multichannel, low nonlinearity time-to-digital
  converters based on 20 and 28 nm FPGAs},'' \emph{IEEE Transactions on
  Industrial Electronics}, vol.~66, no.~4, pp. 3265--3274, 2018.

\bibitem{kalisz1997field}
J.~Kalisz, R.~Szplet, J.~Pasierbinski, and A.~Poniecki,
  ``Field-programmable-gate-array-based time-to-digital converter with 200-ps
  resolution,'' \emph{IEEE Transactions on Instrumentation and Measurement},
  vol.~46, no.~1, pp. 51--55, 1997.

\bibitem{szplet2000interpolating}
R.~Szplet, J.~Kalisz, and R.~Szymanowski, ``{Interpolating time counter with
  100 ps resolution on a single FPGA device},'' \emph{IEEE Transactions on
  Instrumentation and Measurement}, vol.~49, no.~4, pp. 879--883, 2000.

\bibitem{andaloussi2002novel}
M.~S. Andaloussi, M.~Boukadoum, and E.-M. Aboulhamid, ``A novel time-to-digital
  converter with 150 ps time resolution and 2.5 ns pulse-pair resolution,'' in
  \emph{The 14th International Conference on Microelectronics,}, pp.
  123--126.\hskip 1em plus 0.5em minus 0.4em\relax IEEE, 2002.

\bibitem{choi2020design}
K.-J. Choi and D.-W. Jee, ``{Design and Calibration Techniques for a
  Multichannel FPGA-Based Time-to-Digital Converter in an Object Positioning
  System},'' \emph{IEEE Transactions on Instrumentation and Measurement},
  vol.~70, pp. 1--9, 2020.

\bibitem{jarvinen2021100}
O.~J{\"a}rvinen, V.~Unnikrishnan, W.~Siddiqui, T.~Korhonen, K.~Koli,
  K.~Stadius, M.~Kosunen, and J.~Ryyn{\"a}nen, ``{A 100--750 MS/s 11-Bit
  Time-to-Digital Converter With Cyclic-Coupled Ring Oscillator},'' \emph{IEEE
  Access}, vol.~9, pp. 48\,147--48\,156, 2021.

\bibitem{richardson2009low}
J.~A. Richardson, L.~A. Grant, and R.~K. Henderson, ``{Low dark count
  single-photon avalanche diode structure compatible with standard nanometer
  scale CMOS technology},'' \emph{IEEE Photonics Technology Letters}, vol.~21,
  no.~14, pp. 1020--1022, 2009.

\bibitem{tawfeeq2009random}
S.~K. Tawfeeq, ``A random number generator based on single-photon avalanche
  photodiode dark counts,'' \emph{Journal of Lightwave Technology}, vol.~27,
  no.~24, pp. 5665--5667, 2009.

\bibitem{won2016time}
J.~Y. Won and J.~S. Lee, ``{Time-to-digital converter using a tuned-delay line
  evaluated in 28-, 40-, and 45-nm FPGAs},'' \emph{IEEE Transactions on
  Instrumentation and Measurement}, vol.~65, no.~7, pp. 1678--1689, 2016.

\bibitem{pan201420}
W.~Pan, G.~Gong, and J.~Li, ``{A 20-ps time-to-digital converter (TDC)
  implemented in field-programmable gate array (FPGA) with automatic
  temperature correction},'' \emph{IEEE Transactions on Nuclear Science},
  vol.~61, no.~3, pp. 1468--1473, 2014.

\end{thebibliography}

%make it like the style above
%\bibliographystyle{unsrt}
%\bibliography{Reference.bib}
\vspace{-0.5cm}
\begin{IEEEbiography}[{\includegraphics[width=1in,height=1.25in,clip,keepaspectratio]{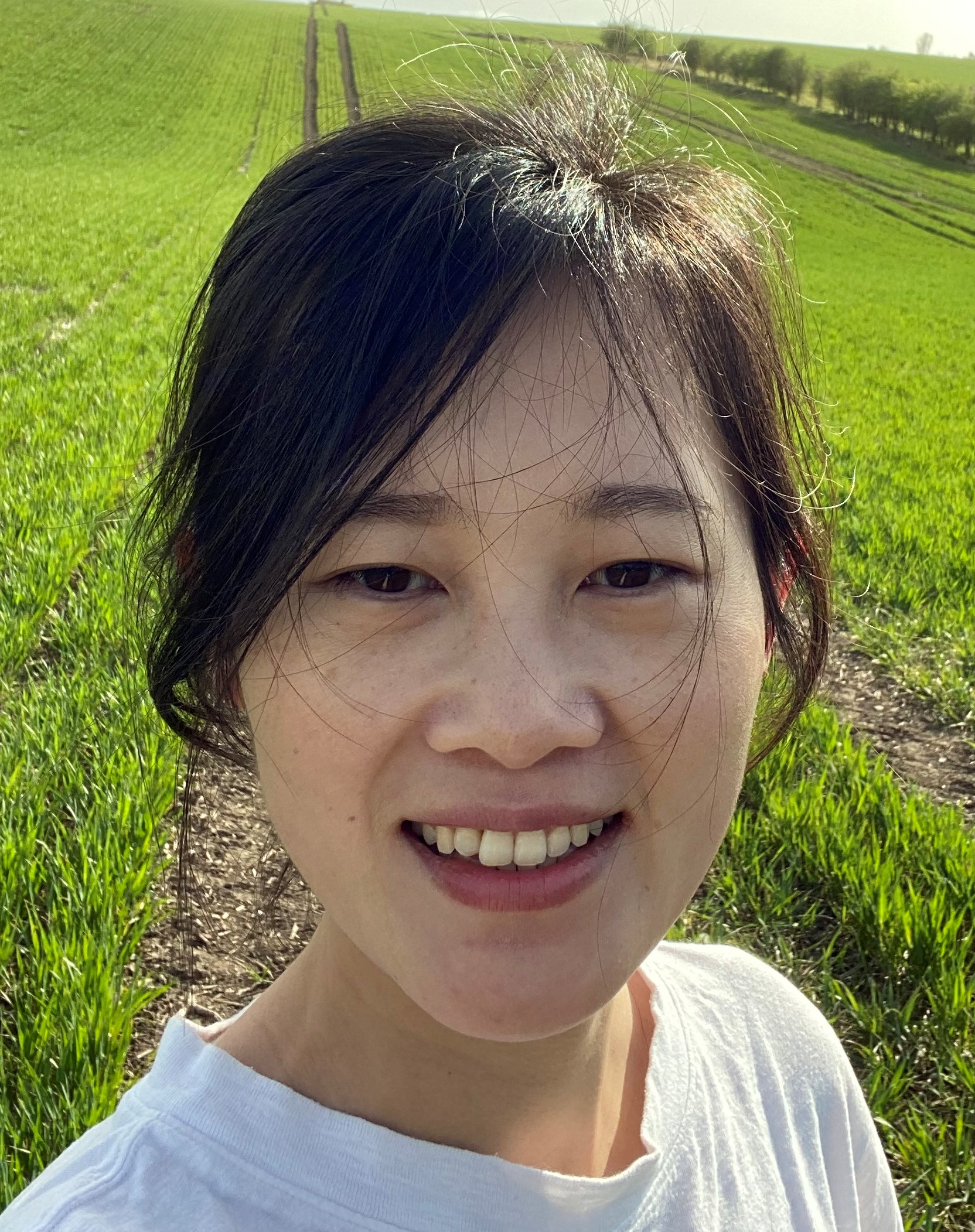}}]
{Yuanyuan Hua} received her MSc in astronomical techniques and methods from Graduate University of Chinese Academy of Science, China in 2011. 
She then joined the Purple Moutain Observatory, Chinese Academy of Science, China, working on high-sensitive and large-field image systems. She is currently pursuing the PhD degree with the University of Edinburgh. Her research interests include image sensors, biomedical imaging, and space imaging.
\end{IEEEbiography}

\vspace{-0.5cm}
\begin{IEEEbiography}[{\includegraphics[width=1in,height=1.25in,clip,keepaspectratio]{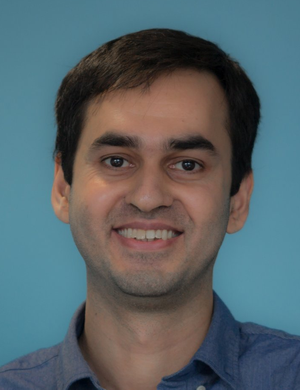}}]
{Danial Chitnis} Danial received his DPhil in engineering science from the University of Oxford in 2013.
He has developed imaging systems including SPAD based image sensors, optical communications, and medical noninvasive brain imaging systems.
SPADs based imaging systems. 
In 2017, he joined the School of Engineering at The University of Edinburgh as a Chancellor's Fellow in Electronics developing detector arrays and systems for a variety of applications from quantum physics to consumer cameras. 

\end{IEEEbiography}
%\nobalance

% References
\end{document}